\documentclass[10pt,conference,a4paper]{IEEEtran}
\usepackage{times,amsmath,epsfig}
\title{Adaptive Set-Membership Reduced-Rank Least Squares Beamforming
Algorithms }
\author{%
{Lei Wang, Rodrigo C. de Lamare}%
\vspace{1.6mm}\\
\fontsize{10}{10}\selectfont
Communications Research Group, Department of Electronics\\
University of York, York YO10 5DD, UK\\
lw517@ohm.york.ac.uk;~~rcdl500@ohm.york.ac.uk}
\begin{document}
\maketitle
\begin{abstract}
This paper presents a new adaptive algorithm for the linearly
constrained minimum variance (LCMV) beamformer design. We
incorporate the set-membership filtering (SMF) mechanism into the
reduced-rank joint iterative optimization (JIO) scheme to develop a
constrained recursive least squares (RLS) based algorithm called
JIO-SM-RLS. The proposed algorithm inherits the positive features of
reduced-rank signal processing techniques to enhance the output
performance, and utilizes the data-selective updates (around
$10-15\%$) of the SMF methodology to save the computational cost
significantly. An effective time-varying bound is imposed on the
array output as a constraint to circumvent the risk of overbounding
or underbounding, and to update the parameters for beamforming. The
updated parameters construct a set of solutions (a membership set)
that satisfy the constraints of the LCMV beamformer. Simulations are
performed to show the superior performance of the proposed algorithm
in terms of the convergence rate and the reduced computational
complexity in comparison with the existing methods.
\end{abstract}

% NOTE keywords are not used for conference papers so do not populate them
% \begin{keywords}
% keyword-1, keyword-2, keyword-3
% \end{keywords}
%
\section{Introduction}
Beamforming is an important antenna array technique with application
in wireless communications to adjust array parameters for
maintaining the array response from a certain direction while
attenuating interference and noise. The optimal linearly constrained
minimum variance (LCMV) beamformer \cite{Frost} is a well-known
beamforming technique. Many adaptive filtering algorithms have been
proposed for the implementation of the LCMV beamformer, ranging from
the low-complexity stochastic gradient (SG) algorithm to the more
complex recursive least squares (RLS) algorithm \cite{Haykin}. The
major drawback of the reported algorithms is that they require a
large number of samples to reach the steady-state when the number of
elements in the filter is large. Besides, filters with many elements
always show a poor performance under dynamic scenarios for tracking
signals embedded in interference and noise.

An effective approach to circumvent these shortcomings is to utilize
reduced-rank array processing and adaptive filtering techniques for
the beamformer design. The reduced-rank array processing techniques
aim to construct a projection matrix to project the array input
vector onto a lower dimensional subspace, and use a reduced-rank
adaptive filter to perform the weight update within this subspace.
Compared with the full-rank techniques, the reduced-rank one
achieves fast convergence and improved tracking performance since
the number of elements in the reduced-rank filter is much less than
those in the full-rank filters, especially under the condition where
the number of sensor elements in the array is large. The popular
reduced-rank schemes include the auxiliary vector filtering (AVF)
\cite{Pados}, the multistage Wiener filter (MSWF) \cite{Honig}, and
the joint iterative optimization (JIO) \cite{Lamare4},
\cite{Lamare}. The SG and RLS type algorithms are developed based on
the reduced-rank schemes for implementation. Despite the improved
convergence and tracking performance achieved with these
reduced-rank adaptive algorithms, the calculation of the projection
matrix and the reduced-rank weight vector requires a significant
computational cost.

The contribution of this paper is the development of a new adaptive
filtering algorithm for the beamformer design that guarantees the
improved convergence and tracking performance compared with those of
the existing full-rank and reduced-rank algorithms, whereas the
computational cost is much lower than that of its reduced-rank
counterparts. An efficient approach to reduce the computational cost
is to employ a set-membership (SM) technique to adaptive filtering
\cite{Guo}, \cite{Diniz}. The SM technique specifies a bound on the
magnitude of the estimation error (or the array output) and uses the
data-selective updates to encompass a set of parameters in a
feasibility set, in which any member is a valid SM filter that
satisfies the constraints of the design criterion. It involves two
steps: $1$) information evaluation and $2$) parameter adaptation. If
the parameter update does not occur frequently, and the information
evaluation does not involve much complexity, the overall
computational cost can be saved substantially. The well-known SM
based algorithms include the works reported in
\cite{Diniz}-\cite{Lamare2}, which were developed for the full-rank
parameter estimation. In this paper, we introduce the SM technique
into reduced-rank array processing and propose a novel reduced-rank
adaptive algorithm. The proposed algorithm introduces a framework to
combine the SM mechanism with the reduced-rank joint iterative
optimization (JIO) scheme \cite{Lamare}, and develops a RLS
algorithm for implementation, which is termed JIO-SM-RLS. An
effective time varying bound is employed in the proposed algorithm
as a constraint to avoid the risk of overbounding or underbounding
\cite{Lamare2}. Compared with the existing algorithms, the
JIO-SM-RLS algorithm inherits the positive features of the JIO
scheme to enhance the convergence and tracking performance, and
utilizes the data-selective updates of the SM mechanism to save the
computational cost significantly.

The remaining of this paper is organized as follows: we outline a
system model for beamforming and present the reduced-rank technique
in Section II. Section III introduces the reduced-rank SM scheme and
Section IV derives the proposed JIO-SM-RLS algorithm. Simulation
results are provided and discussed in Section V, and conclusions are
drawn in Section VI.

\section{System Model and Reduced-rank\\ Beamformer Design}

\subsection{System Model}

Let us suppose that $q$ narrowband signals impinge on a uniform
linear array (ULA) of $m$ ($q \leq m$) sensor elements. The sources
are assumed to be in the far field with directions of arrival (DOAs)
$\theta_{0}$,\ldots,$\theta_{q-1}$. The received vector $\boldsymbol
x(i)\in\mathcal C^{m\times 1}$ at the $i$th snapshot can be modeled
as

\begin{equation} \label{1}
\centering {\boldsymbol x}(i)={\boldsymbol A}({\boldsymbol
{\theta}}){\boldsymbol s}(i)+{\boldsymbol n}(i),~~~ i=1,\ldots,N
\end{equation}
where
$\boldsymbol{\theta}=[\theta_{0},\ldots,\theta_{q-1}]^{T}\in\mathcal{R}^{q
\times 1}$ is the DOAs, ${\boldsymbol A}({\boldsymbol
{\theta}})=[{\boldsymbol a}(\theta_{0}),\ldots,{\boldsymbol
a}(\theta_{q-1})]\in\mathcal{C}^{m \times q}$ composes the steering
vectors ${\boldsymbol a}(\theta_{k})=[1,e^{-2\pi
j\frac{d}{\lambda_{c}}cos{\theta_{k}}},\ldots,e^{-2\pi
j(m-1)\frac{d}{\lambda_{c}}cos{\theta_{k}}}]^{T}\in\mathcal{C}^{m
\times 1},~~~(k=0,\ldots,q-1)$, where $\lambda_{c}$ is the
wavelength and $d=\lambda_{c}/2$ is the inter-element distance of
the ULA, and to avoid mathematical ambiguities, the steering vectors
$\boldsymbol a(\theta_{k})$ are considered to be linearly
independent, ${\boldsymbol s}(i)\in \mathcal{C}^{q\times 1}$ is the
source data, ${\boldsymbol n}(i)\in\mathcal{C}^{m\times 1}$ is the
white Gaussian noise, $N$ is the observation size of snapshots, and
$(\cdot)^{T}$\ stands for the transpose. The output of a narrowband
beamformer is

\begin{equation} \label{2}
\centering y(i)={\boldsymbol w}^H{\boldsymbol x}(i),
\end{equation}
where ${\boldsymbol
w}=[w_{1},\ldots,w_{m}]^{T}\in\mathcal{C}^{m\times 1}$ is the
complex weight vector of the adaptive filter, and $(\cdot)^{H}$
stands for the Hermitian transpose.

\subsection{Reduced-rank Beamformer Design}

For large $m$ or in the dynamic scenario, the full-rank adaptive
algorithms (e.g., SG or RLS) fail or provide poor performance with a
small number of snapshots  for the beamformer design. Many of recent
works in the literature have been reported based on the reduced-rank
techniques to solve these problems \cite{Pados}-\cite{Lamare4}. The
important feature of the reduced-rank schemes is to construct a
projection matrix $\boldsymbol T_r=[\boldsymbol t_1, \ldots,
\boldsymbol t_r]\in\mathcal C^{m\times r}$ with columns $\boldsymbol
t_l$ ($l=1, \ldots, r$) constitute a bank of $r$ full-rank filters
as given by $\boldsymbol t_l=[t_{1,l}, \ldots, t_{m,l}]^T\in\mathcal
C^{m\times1}$. The projection matrix performs the dimensionality
reduction to project the full-rank received-vector onto a lower
dimension and retains the key information of the original signal in
a reduced-rank received vector, which is

\begin{equation}\label{3}
\bar{\boldsymbol x}(i)=\boldsymbol T_r^H\boldsymbol x(i),
\end{equation}
where $\bar{\boldsymbol x}\in\mathcal C^{r\times1}$ denotes the
reduced-rank received vector and $r$ ($1\leq r\leq m$) is the rank
number. In what follows, all $r$-dimensional quantities are denoted
by an over bar.

The reduced-rank adaptive filter $\bar{\boldsymbol w}=[\bar{w}_1,
\ldots, \bar{w}_r]$ follows the projection matrix to produce the
filter output

\begin{equation}\label{4}
y(i)=\bar{\boldsymbol w}^H\bar{\boldsymbol x}(i).
\end{equation}

The popular reduced-rank schemes include the AVF \cite{Pados}, the
MSWF \cite{Honig}, and the JIO \cite{Lamare}, which employ the SG or
the RLS type algorithms to calculate $\boldsymbol T_r$ and
$\bar{\boldsymbol w}$ for the beamformer design. However, with
respect to the SG type algorithms, it is difficult to predetermine
the step size values to make a tradeoff between fast convergence and
misadjustment in dynamic scenarios. Furthermore, the computational
cost is high due to the calculation of the projection matrix.

\section{Proposed Reduced-rank SM Scheme}

In this section, we introduce a novel reduced-rank SM scheme by
combining the SM mechanism with the reduced-rank JIO scheme. It
should be remarked that the JIO scheme is considered here since it
exhibits the superior convergence and tracking performance with
relatively simple realization over other reduced-rank schemes
\cite{Lamare}.

The existing SM techniques focus on full-rank signal processing,
namely, the related filter $\boldsymbol w$ is encompassed in a
feasibility set, in which any member satisfies a predetermined or
time-varying bound on the magnitude of the estimation error (or the
array output). The SM algorithms utilize the data-selective updates
to reduce computational complexity \cite{Lamare2}. Regarding the
proposed reduced-rank SM scheme, some valid pairs of $\{\boldsymbol
T_r, \bar{\boldsymbol w}\}$ are consistent with the bound at each
time instant due to the joint iterative exchange of information. The
solution to the proposed scheme is a feasibility set in the
parameter space, which is

\begin{small}
\begin{equation}\label{5}
\Theta(i)=\bigcap_{\big(s_0(i),\boldsymbol x(i)\big)\in\mathcal
{\boldsymbol S}}\Big\{\bar{\boldsymbol w}\in\mathcal C^{r\times
1},\boldsymbol T_r\in\mathcal C^{m\times
r}:|y(i)|^2\leq\delta^2(i)\Big\},
\end{equation}
\end{small}
where $s_0(i)$ is the transmitted data of the desired user from
$\theta_0$ and $\boldsymbol S$ is the set of all possible data pairs
$(s_0(i), \boldsymbol x(i))$. The pairs of $\{\boldsymbol T_r,
\bar{\boldsymbol w}\}$ in the set are upper bounded in magnitude by
a time-varying bound $\delta(i)$ that can be viewed as a constrained
condition in the beamformer design. Actually, $\boldsymbol S$ cannot
be traversed all over in practice. An alternative way is to
construct an exact membership set $\Psi(i)$, which is the
intersection of the constraint sets provided by the observations
over the time instants $i=1, \ldots, N$, i.e.,
$\Psi(i)=\cap_{l=1}^i\mathcal {H}_l$ with the constraint set
$\mathcal H_l=\{\bar{\boldsymbol w}\in\mathcal C^{r\times 1},
\boldsymbol T_r\in\mathcal C^{m\times r}:
|y(i)|^2\leq\delta^2(i)\}$. It is clear that a larger space of the
data pairs leads to a smaller membership set. Note that the
feasibility set $\Theta(i)$ is a subset of the exact membership set
$\Psi(i)$. The two sets will be equal if the data pairs traverse
$\boldsymbol S$ completely.

\section{Proposed JIO-SM-RLS Algorithm}

In this section, we employ the proposed reduced-rank SM scheme to
develop a new RLS algorithm. The objective is to design a bank of
full-rank filters and a reduced-rank filter whose output is not
greater than a time-varying bound but remains the signal from one
certain direction for all input data. It can be derived by
minimizing the following cost function

\begin{equation}\label{6}
\begin{split}
&\textrm{minimize}~~E\big[|\bar{\boldsymbol
w}^H\boldsymbol T_r^H{\boldsymbol x}(i)|^2\big]\\
&\textrm{subject~to}~~\bar{\boldsymbol w}^H\bar{\boldsymbol
a}(\theta_0)=\gamma,~~\textrm{and}~~|\bar{\boldsymbol
w}^H\boldsymbol T_r^H{\boldsymbol x}(i)|^2=g^2(i),
\end{split}
\end{equation}
where $\boldsymbol a(\theta_0)$ and $\bar{\boldsymbol a}(\theta_0)$
are the full-rank and the reduced-rank steering vectors of the
desired signal, $\gamma$ is a constant with respect to the
constraint, $\bar{\boldsymbol x}(i)=\boldsymbol T_r^H\boldsymbol
x(i)$ is the reduced-rank received vector, and $g(i)$ corresponds to
a bound within the constraint set $\mathcal {H}_i$ with
$|g(i)|\leq\delta(i)$. The solutions $\{\boldsymbol T_r,
\bar{\boldsymbol w}\}$ construct the feasibility set in (\ref{5})
and satisfy the constraints in (\ref{6}).

The constrained cost function can be transferred into an
unconstrained least squares (LS) cost function by using the method
of Lagrange multipliers \cite{Haykin}, which is

\begin{equation}\label{7}
\begin{split}
&J_{\textrm{LS}}=\sum_{l=1}^{i-1}\lambda_1^{i-l}(i)\bar{\boldsymbol
w}^H(i)\boldsymbol T_r^H(i)\boldsymbol x(l)\boldsymbol
x^H(l)\boldsymbol T_r(i)\bar{\boldsymbol w}(i)\\
&+\lambda_1(i)\big[|\bar{\boldsymbol w}^H(i)\boldsymbol
T_r^H(i)\boldsymbol
x(i)|^2-g^2(i)\big]+\lambda_2\big[\bar{\boldsymbol
w}^H(i)\bar{\boldsymbol a}(\theta_0)-\gamma\big],
\end{split}
\end{equation}
where $\lambda_1(i)$ plays the role of the forgetting factor and
Lagrange multiplier with respect to the constraint on the amplitude
of the array output, and $\lambda_2$ denotes another Lagrange
multiplier for the constraint on the steering vector.

It is clear that (\ref{7}) is a function of the projection matrix
$\boldsymbol T_r(i)$ and the reduced-rank filter $\bar{\boldsymbol
w}(i)$. Taking the gradient of $\bar{\boldsymbol w}(i)$ with respect
to (\ref{7}) and making it equal to a null vector, we have

\begin{small}
\begin{equation}\label{8}
\begin{split}
\bar{\boldsymbol
w}(i)=&-\lambda_2\Big\{\big[\sum_{l=1}^{i-2}\lambda_1^{i-l}(i)\bar{\boldsymbol
x}(l)\bar{\boldsymbol x}^H(l)+\lambda_1(i)\bar{\boldsymbol
x}(i-1)\bar{\boldsymbol
x}^H(i-1)\big]\\
&+\lambda_1(i)\bar{\boldsymbol x}(i)\bar{\boldsymbol
x}^H(i)\Big\}^{-1}\bar{\boldsymbol a}(\theta_0)\\
&\approx-\lambda_2\bar{\boldsymbol R}^{-1}(i)\bar{\boldsymbol
a}(\theta_0),
\end{split}
\end{equation}
\end{small}
where $\bar{\boldsymbol R}(i)=\bar{\boldsymbol
R}(i-1)+\lambda_1(i)\bar{\boldsymbol x}(i)\bar{\boldsymbol x}^H(i)$.
It should be remarked that the second expression of (\ref{8}) is
obtained under an assumption that $\lambda_1(i)\rightarrow1$, which
is in accordance with the setting of the forgetting factor
\cite{Haykin}.

By substituting (\ref{8}) into the first constraint in (\ref{6}) and
employing the matrix inversion lemma \cite{Haykin} to solve
$\lambda_2$, we get

\begin{equation}\label{9}
\bar{\boldsymbol w}(i)=\frac{\gamma\bar{\boldsymbol
P}(i)\bar{\boldsymbol a}(\theta_0)}{\bar{\boldsymbol
a}^H(\theta_0)\bar{\boldsymbol P}(i)\bar{\boldsymbol a}(\theta_0)},
\end{equation}
where $\bar{\boldsymbol P}(i)=\bar{\boldsymbol R}^{-1}(i)$ is
calculated in a recursive form

\begin{equation}\label{10}
\bar{\boldsymbol k}(i)=\frac{\bar{\boldsymbol
P}(i-1)\bar{\boldsymbol x}(i)}{1+\lambda_1(i)\bar{\boldsymbol
x}^H(i)\bar{\boldsymbol P}(i-1)\bar{\boldsymbol x}(i)}
\end{equation}
\begin{equation}\label{11}
\bar{\boldsymbol P}(i)=\bar{\boldsymbol
P}(i-1)-\lambda_1(i)\bar{\boldsymbol k}(i-1)\bar{\boldsymbol
x}^H(i)\bar{\boldsymbol P}(i-1).
\end{equation}

Taking the gradient of $\boldsymbol T_r(i)$ with respect to
(\ref{7}), making it equal to a zero matrix, and consider the
assumption $\lambda_1(i)\rightarrow1$, we obtain

\begin{equation}\label{12}
\boldsymbol T_r(i)\bar{\boldsymbol w}(i)=-\lambda_2\boldsymbol
R^{-1}(i)\boldsymbol a(\theta_0),
\end{equation}
where $\boldsymbol R(i)=\boldsymbol R(i-1)+\lambda_1(i)\boldsymbol
x(i)\boldsymbol x^H(i)$. If we define $\boldsymbol f(i)=\boldsymbol
R^{-1}(i)\boldsymbol a(\theta_0)$, the solution of $\boldsymbol
T_r(i)$ in (\ref{12}) can be regarded to find the solution to the
linear equation $\boldsymbol T_r(i)\bar{\boldsymbol
w}(i)=\boldsymbol f(i)$. Given a $\bar{\boldsymbol
w}(i)\neq\boldsymbol 0$, there exists multiple $\boldsymbol T_r(i)$
in general. We derive the minimum Frobenius-norm solution for
stability. The details of this derivation can be found in
\cite{Wang}. The projection matrix can be expressed by

\begin{equation}\label{13}
\boldsymbol T_r(i)=-\lambda_2\boldsymbol R^{-1}(i)\boldsymbol
a(\theta_0)\frac{\bar{\boldsymbol w}^H(i)}{\|\bar{\boldsymbol
w}(i)\|^2}.
\end{equation}

The Lagrange multiplier $\lambda_2$ can be solved by substituting
(\ref{13}) into the constraint $\bar{\boldsymbol w}^H(i)\boldsymbol
T_r^H(i)\boldsymbol a(\theta_0)=\gamma$. After several
rearrangements, the resultant projection matrix becomes

\begin{equation}\label{14}
\boldsymbol T_r(i)=\frac{\gamma\boldsymbol P(i)\boldsymbol
a(\theta_0)}{\boldsymbol a^H(\theta_0)\boldsymbol P(i)\boldsymbol
a(\theta_0)}\frac{\bar{\boldsymbol w}^H(i)}{\|\bar{\boldsymbol
w}(i)\|^2},
\end{equation}
where $\boldsymbol P(i)=\boldsymbol R^{-1}(i)$ is calculated by

\begin{equation}\label{15}
\boldsymbol k(i)=\frac{\boldsymbol P(i-1)\boldsymbol
x(i)}{1+\lambda_1(i)\boldsymbol x^H(i)\boldsymbol P(i-1)\boldsymbol
x(i)}
\end{equation}
\begin{equation}\label{16}
\boldsymbol P(i)=\boldsymbol P(i-1)-\lambda_1(i)\boldsymbol
k(i)\boldsymbol x^H(i)\boldsymbol P(i-1).
\end{equation}

The coefficient $\lambda_1(i)$ is important to the updates of the
projection matrix $\boldsymbol T_r(i)$ and the reduced-rank filter
$\bar{\boldsymbol w}(i)$. It guarantees an effective exchange of
information between $\boldsymbol T_r(i)$ and $\bar{\boldsymbol
w}(i)$, and keeps the constraint on the amplitude of the array
output upper bounding a specific value following the time instant.
We utilize the proposed reduced-rank SM scheme to compute
$\lambda_1(i)$ and perform data-selective updates to adjust pairs of
$\{\boldsymbol T_r(i), \bar{\boldsymbol w}(i)\}$ with low
complexity. Specifically, substituting the expressions of (\ref{9})
and (\ref{14}) into the second constraint in (\ref{6}) and making a
rearrangement, yields,

%\begin{small}
\begin{equation}\label{17}
\begin{split}
&\lambda_1(i)=\\
&\left\{ \begin{array}{ccc}
                  \frac{\boldsymbol a^H(\theta_0)\boldsymbol
P(i-1)[\delta(i)\boldsymbol a(\theta_0)-\gamma^2\boldsymbol
x(i)]}{\boldsymbol a^H(\theta_0)\boldsymbol k(i)\boldsymbol
x^H(i)\boldsymbol P(i-1)[\delta(i)\boldsymbol
a(\theta_0)-\gamma^2\boldsymbol x(i)]}
                    & \textrm{if} ~|y(i)|^2\geq\delta^2(i)\\
                  0 & \textrm {otherwise},\\
                  \end{array}\right.
\end{split}
\end{equation}
%\end{small}
where $\boldsymbol k(i)$ has been given in (\ref{15}). The
coefficient $\lambda_1(i)$ is calculated only if the constraint
$|\bar{\boldsymbol w}^H(i)\bar{\boldsymbol x}(i)|^2=g^2(i)$ cannot
be satisfied, so as the updates of $\boldsymbol T_r(i)$ and
$\bar{\boldsymbol w}(i)$. It provides the data-selective updates for
the full-rank and reduced-rank filters' design, reduces the
computational complexity significantly, and encompasses pairs of
$\{\boldsymbol T_r(i), \bar{\boldsymbol w}(i)\}$ in the feasibility
set $\Theta(i)$ proposed in Section 3.

In (\ref{17}), $\lambda_1(i)$ is sensitive to the selection of the
time-varying bound $\delta(i)$, which impacts the update rate and
the tracking performance of the proposed algorithm. We describe a
parameter dependent bound (PDB) that is similar to the work reported
in \cite{Guo}, and considers the evolution of the full-rank weight
vector $\boldsymbol w(i)=\boldsymbol T_r(i)\bar{\boldsymbol w}(i)$
to make the proposed algorithm work effectively. The proposed
time-varying bound is

\begin{equation}\label{18}
\delta(i)=\beta\delta(i-1)+(1-\beta)\sqrt{\alpha\|\boldsymbol
T_r(i)\bar{\boldsymbol w}(i)\|^2\hat{\sigma}_n^2(i)},
\end{equation}
where $\beta$ is a forgetting factor that should be set to guarantee
an appropriate time-averaged estimate of the evolution of the weight
vector $\boldsymbol w(i)$, $\alpha$ ($\alpha>1$) is a tuning
coefficient, $\hat{\sigma}_n^2(i)$ is an estimate of the noise
power, and $\|\boldsymbol T_r(i)\bar{\boldsymbol
w}(i)\|^2\hat{\sigma}_n^2(i)$ is the variance of the inner product
of the weight vector with the noise term $\boldsymbol n(i)$ that
provides information on the evolution of $\boldsymbol w(i)$. The
proposed time-varying bound provides a smoother evolution of the
weight vector trajectory and thus avoids too high or low values of
the squared norm of the weight vector.

A summary of the proposed JIO-SM-RLS algorithm is given in Table
\ref{tab:JIO-SM-RLS}, where $\rho$ and $\varrho$ are small positive
values for regularization, and $\boldsymbol T_r(0)$ and
$\bar{\boldsymbol w}(0)$ are given to ensure the constrained
condition. It is clear that the projection matrix and the
reduced-rank filter exchange information and rely on each other,
which leads to an improved convergence and tracking performance for
the proposed algorithm. The proposed reduced-rank SM scheme with the
time-varying bound is employed in the devised algorithm to update
the pairs of $\{\boldsymbol T_r(i), \bar{\boldsymbol w}(i)\}$ only
when the constraint on the array output power cannot be satisfied,
which results in substantial savings in computation that is much
less than that of its conventional counterparts.

\begin{table}[!t]
\centering
    \caption{THE PROPOSED JIO-SM-RLS ALGORITHM}     % NOTE!  caption goes _before_ the table contents !!
    \label{tab:JIO-SM-RLS}
    \begin{small}
        \begin{tabular}{|l|}
\hline
\bfseries {Initialization:}\\
\hline
~~~~~~~${\boldsymbol T}_r(0)=[{\boldsymbol I}_{r\times r}~\boldsymbol 0_{r\times (m-r)}]^T$\\
~~~~~~~${\bar{\boldsymbol w}}(0)=\boldsymbol T_r^H(0)\boldsymbol a(\theta_0)/(\|\boldsymbol T_r^H(0)\boldsymbol a(\theta_0)\|^2)$\\
~~~~~~~$\boldsymbol P(0)=\rho\boldsymbol I_{m\times m}$\\
~~~~~~~$\bar{\boldsymbol P}(0)=\varrho\boldsymbol I_{r\times r}$\\
\hline
\bfseries {For each time instant} $i=1,\ldots, N$\\
\hline
~~~~~~~$\bar{\boldsymbol x}(i)=\boldsymbol
T_r^H(i-1)\boldsymbol
x(i)$\\
~~~~~~~$y(i)=\bar{\boldsymbol w}^H(i-1)\bar{\boldsymbol x}(i)$\\
~~~~~~~$\delta(i)$~~~\textrm{in}~~~~~~~~~~~~~~~~~~~~~~~~~~~~~~~~~~~~~~~~~~(\ref{18})\\
~~~~~~~\bfseries {if} ~~~$|y(i)|^2\geq\delta^2(i)$\\
~~~~~~~~~~~~~$\lambda_1(i)$~~~~\textrm{in}~~~~~~~~~~~~~~~~~~~~~~~~~~~~~~~~~~(\ref{17})\\
~~~~~~~~~~~~~$\boldsymbol k(i)$~~~~~\textrm{in}~~~~~~~~~~~~~~~~~~~~~~~~~~~~~~~~~~(\ref{15})\\
~~~~~~~~~~~~~$\boldsymbol
P(i)$~~~~~\textrm{in}~~~~~~~~~~~~~~~~~~~~~~~~~~~~~~~~~(\ref{16})\\
~~~~~~~~~~~~~$\boldsymbol T_r(i)$~~~~\textrm{in}~~~~~~~~~~~~~~~~~~~~~~~~~~~~~~~~~(\ref{14})\\
~~~~~~~~~~~~~$\bar{\boldsymbol a}(\theta_0)=\boldsymbol T_r^H(i)\boldsymbol a(\theta_0)$\\
~~~~~~~~~~~~~$\boldsymbol x(i)=\boldsymbol T_r^H(i)\boldsymbol x(i)$\\
~~~~~~~~~~~~~$\bar{\boldsymbol k}(i)$~~~~~~\textrm{in}~~~~~~~~~~~~~~~~~~~~~~~~~~~~~~~~~(\ref{10})\\
~~~~~~~~~~~~~$\bar{\boldsymbol P}(i)$~~~~~\textrm{in}~~~~~~~~~~~~~~~~~~~~~~~~~~~~~~~~~(\ref{11})\\
~~~~~~~~~~~~~$\bar{\boldsymbol w}(i)$~~~~~\textrm{in}~~~~~~~~~~~~~~~~~~~~~~~~~~~~~~~~~~(\ref{9})\\
~~~~~~\bfseries {else}\\
~~~~~~~~~~~~~$\boldsymbol T_r(i)=\boldsymbol T_r(i-1)$\\
~~~~~~~~~~~~~$\bar{\boldsymbol w}(i)=\bar{\boldsymbol w}(i-1)$\\
~~~~~~\bfseries {end}\\
\hline
    \end{tabular}
    \end{small}
\end{table}

\section{Simulation Results}

In this section, we evaluate the output signal-to-interference
plus-noise ratio (SINR) performance of the proposed JIO-SM-RLS
algorithm and compare it with the existing methods, including the
full-rank (FR) SG and RLS type algorithms with or without SM
techniques \cite{Haykin}, \cite{Lamare2}, and reduced-rank
algorithms based on the AVF \cite{Pados}, the MSWF \cite{Honig}, and
the JIO \cite{Lamare} schemes. We assume that the DOA of the desired
user is known by the receiver. In each experiment, we consider BPSK
signals and set input SNR$=10$ dB and INR$=30$ dB with white
Gaussian noise. Simulations are carried out with a ULA containing
$m=64$ sensor elements with half-wavelength interelement spacing. A
total of $K=1000$ runs are performed to obtain each curves.

In the first experiment, $q=25$ users, including one desired user,
exist in the system. The related coefficients for the proposed
algorithm are set $\gamma=1$, $r=5$, $\alpha=26$, $\beta=0.992$,
$\rho=1.3\times10^{-3}$, and $\varrho=1.0\times10^{-4}$. It should
be remarked that $\lambda_1(i)$ should be in accordance with the
setting of the forgetting factor, which is a small positive value
less than $1$. In simulations, we limit its range
$0.1\leq\lambda_1(i)\leq0.998$ for implementation. In Fig. 1, the
curve of the proposed JIO-SM-RLS algorithm achieves superior
convergence compared with existing ones. The steady-state
performance of the proposed algorithm is quite close to that of the
minimum variance distortionless response (MVDR) that assumes the
knowledge of the covariance matrix $\boldsymbol R$ \cite{Haykin}.
Although the JIO-RLS algorithm \cite{Lamare} also enjoys relatively
good performance, it requires $100\%$ updates ($1000$ updates for
$1000$ snapshots) for the filter design, which is quite higher than
that of the proposed algorithm with only $14.2\%$ updates for the
pairs of $\{\boldsymbol T_r(i), \bar{\boldsymbol w}(i)\}$.

\begin{figure}[h]
    \centerline{\psfig{figure=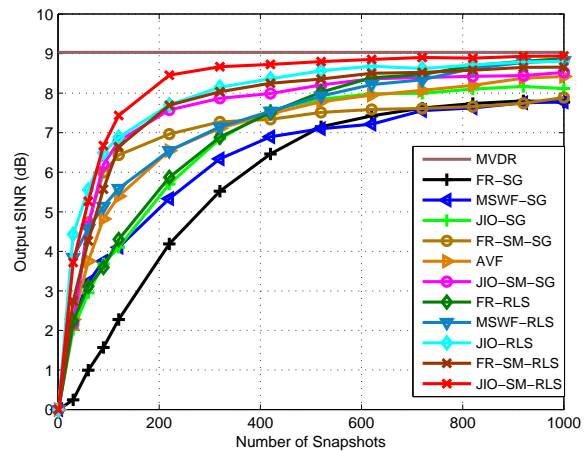,height=64.54mm} }
    \caption{Output SINR versus the
number of snapshots}
    \label{fig:all_methods}
\end{figure}

The next simulation includes two experiments, which compare the
proposed and existing algorithms with the time-varying and fixed
bounds, respectively. The scenario is the same as that in Fig. 1.
Fig. 2 (a) shows the results for the full-rank algorithms. We find
that the FR-SM-RLS algorithm converges quickly to the steady-state
with relatively low update rate ($\tau=20.5\%$). Due to the BPSK
modulation scheme, it implies that the algorithm with the fixed
bound $\delta=1.0$ should achieve a good performance. However, it
requires more updates ($\tau=61.4\%$) and thus increases the
computational cost. The curves with higher ($\delta=1.4$) or lower
($\delta=0.8$) bounds exhibit worse convergence performance. The
same result can be found in Fig. 2 (b) for the reduced-rank
algorithms. The proposed algorithm with the time-varying bound uses
even less updates to realize a high output SINR performance.

\begin{figure}[h]
    \centerline{\psfig{figure=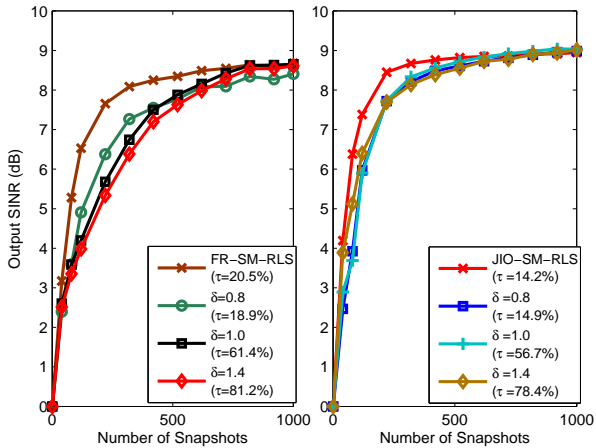,height=64.54mm} }
    \caption{Output SINR versus the
number of snapshots for (a) full-rank SM algorithms; (b)
reduced-rank SM algorithms}
    \label{fig:sm_fix}
\end{figure}

\section{Conclusion}

We have introduced a new reduced-rank SM scheme and develop a RLS
algorithm for implementation. The proposed scheme incorporates the
SM mechanism with the time-varying bound into the reduced-rank JIO
scheme to realize the data-selective updates of the full-rank and
reduced-rank filters. The time-varying bound is combined in the LCMV
optimization problem as a new constraint on the array output power
to encompass the pairs of $\{\boldsymbol T_r(i), \bar{\boldsymbol
w}(i)\}$ in the feasibility set of the proposed scheme. A RLS
algorithm has been derived for implementation. The proposed
algorithm achieves an improved performance with exchange of
information between the projection matrix and the reduced-rank
weight vector, reducing the computational cost primarily with the
data-selective updates.

\end{document}